\documentclass[aps,prd,onecolumn,amsmath,showpacs,showkeys, superscriptaddress,nofootinbib]{revtex4}

\usepackage[colorlinks,citecolor=blue,urlcolor=blue,linkcolor=blue]{hyperref}
\usepackage{array,multirow,graphicx}
\newlength{\extraspace}
\newlength{\extraspaces}
\begingroup
\endgroup

\newcommand{\be}{\begin{equation}
\addtolength{\abovedisplayskip}{\extraspaces}
\addtolength{\belowdisplayskip}{\extraspaces}
\addtolength{\abovedisplayshortskip}{\extraspace}
\addtolength{\belowdisplayshortskip}{\extraspace}}
\newcommand{\ee}{\end{equation}}

\newcommand{\ba}{\begin{eqnarray}
\addtolength{\abovedisplayskip}{\extraspaces}
\addtolength{\belowdisplayskip}{\extraspaces}
\addtolength{\abovedisplayshortskip}{\extraspace}
\addtolength{\belowdisplayshortskip}{\extraspace}}
\newcommand{\ea}{\end{eqnarray}}

\newcommand{\nonu}{\nonumber \\[.5mm]}
\newcommand{\A}{&\!\!\!}

\newcommand{\newsection}[1]{
\vspace{7mm} \pagebreak[3] \addtocounter{section}{1}
\setcounter{subsection}{0} 
\begin{center}
{\large {\bf \thesection. #1}}
\end{center}
\nopagebreak
\medskip
\nopagebreak \hspace{3mm}}

\begin{document}

\title{  Magnetic black holes in Weitzenb\"ock geometry}


\author{ Gamal  G. L. Nashed }
\affiliation{Centre for Theoretical Physics, The British University in Egypt, P.O. Box 43, El Sherouk City, Cairo 11837, Egypt.}
\affiliation{Department of Mathematics, Faculty of Science, Ain Shams University, Cairo 11566, Egypt.}

\author{Salvatore Capozziello}
\affiliation{Dipartimento di Fisica ``E. Pancini``,  Universit\`a di Napoli ``Federico II'',
Complesso Universitario di Monte Sant' Angelo, Edificio G, Via Cinthia, I-80126, Napoli, Italy.}
\affiliation{Istituto Nazionale di Fisica Nucleare (INFN),  Sezione di Napoli,
Complesso Universitario di Monte Sant'Angelo, Edificio G, Via Cinthia, I-80126, Napoli, Italy.}
\affiliation{Laboratory for Theoretical Cosmology,
Tomsk State University of Control Systems and Radioelectronics (TUSUR), 634050 Tomsk, Russia.
}
\begin{abstract}
 We derive   magnetic black hole solutions  using a general gauge potential in the framework of teleparallel equivalent general relativity. One of the solutions gives a non-trivial value of the scalar torsion. This non-triviality of the torsion scalar depends on some values of the magnetic field. The metric   of those solutions behave asymptotically as Anti-de-Sitter/ de-Sitter (AdS/dS) spacetimes. The energy conditions are discussed in details.  Also, we calculate the  torsion and curvature invariants to discuss  singularities.  Additionally, we calculate the conserved quantities  using the  Einstein-Cartan geometry to understand  the physics of the constants appearing  into  the solutions.
\keywords{ Teleparallel gravity; black hole solutions; singularities, conserved quantities.}
\pacs{ 04.50.Kd, 98.80.-k, 04.80.Cc, 95.10.Ce, 96.30.-t}
\end{abstract}
\maketitle

\begin{center}
\newsection{\bf Introduction}
\end{center}
Several issues  related to the gravitational field, ranging  from   quantum gravity up to cosmological    dark energy and dark matter,  encourage scientists to search for modifications of general relativity (GR)  capable of addressing under a new standard the phenomenology \cite{rept,starx, Nojiri, CFPS,repva,cai,sebastiani}. Clearly, a viable modification must be  consistent  with  experimental tests and solve   problems in quantum gravity and/or cosmology, that is at UV and IR scales.  From a theoretical point of view,  a first requirement is avoiding  ghosts or other severe defects  to achieve self-consistent theories \cite{GKS}.

Therefore, it is logic  to start using different approaches with respect to  GR. Among these constructions, there  is the one   used by Einstein himself \cite{Ea}-\cite{Ea2}. Assuming absolute parallelism, we can  formulate  gravitational field theory equivalent to GR using tetrad fields as  building blocks instead of the metric \cite{M2}-\cite{Obukhov}.  In this theory the gravitational field is due to  torsion tensor, which is acting as a force \cite{AP97},  instead of  the curvature tensor of GR. This theory is known in the literature as   teleparallel equivalent of general relativity (TEGR)  which  is equivalent to GR and  allows  new conjectures into gravitation \cite{Pj12}. The main  advantage of this theory  is that one can define a true  gravitational energy-momentum tensor  which is  locally  Lorentz  invariant \cite{Mj9,MRTC,capriolo}. Generally, we can  consider TEGR as a gauge theory of translation \cite{Cy}-\cite{Tr}. In this  framework,  the tetrad field  has the role of the  gauge translational potential  of gravitational field \cite{ORP}.

It is worth stressing that Einstein himself used the Weitzenb\"ock geometry with the aim to unify the electromagnetic and gravitational fields \cite{Ea2}. The issue is connected to  the fact that  the tetrad field has 16-components while  the metric field has  10-components. The 16-components of the tetrad can be regarded as 10-components encoded for the metric and  6-components as  degrees of freedom   responsible for the electromagnetic field. However, the  possible unification pointed out  by Einstein  failed but the concept to adopt the Weitzenb\"ock geometry and  TEGR to describe the gravitational field through the torsion field remained. In particular, considering    tetrads as   fields in the context of TEGR, many authors  investigated the notion of gravitational energy in different physical cases \cite{M13,M10,M07}. Specifically, the notion of gravitational energy has been adopted to construct a non-local gravitational theory \cite{M15}.

 A crucial issue in both GR and TEGR is to derive exact black hole solutions in order to understand the main  features of the theory. An important topic is  to derive asymptotically flat black hole solutions in the context of Einstein-Maxwell   equations of motions, either rotating or non-rotating. The applications of such solutions to  stellar systems is straightforward: in particular, if these systems are  endowed with a magnetic field they can represent realistic astrophysical objects.  In fact many charged black hole solutions have been derived using Einstein-Maxwell field equations and  not all of them,  in  static situations,  coincide with the Schwarzschild spacetime \cite{B66}-\cite{B61}.

Furthermore, many solutions have been derived in  TEGR theory \cite{Nepjc07,NSijmpd07}, however,  till now,  no magnetic black holes  with flat horizons  have been derived. A flat horizon is  a spacetime with cylindrical symmetry. This spacetime  plays a main  role in the discussion
of  internal consistency of a given solution as in the case of   Levi-Civita \cite{Ct,Wh} and Chazy-Curzon  \cite{Cj,Ch} static solutions, and the
 Lewis solutions \cite{Lt}.  In astrophysical context, cylindrical symmetry has been used
to  study  cosmic strings \cite{Va}. In  GR, for example,  cylindrically symmetric rotating black hole solutions have been derived with a negative cosmological constant \cite{Lj,Aa}. The aim of the present study is  to derive  magnetic black holes using general gauge potential in the framework of TEGR and analyze their physical properties   by discussing  their energy conditions, studying  their singularities and  calculating the related  conserved charges.

The layout  of the paper is the  following: In \S II, a summary TEGR geometry is provided. In \S III, a tetrad field with  cylindrical  symmetry  is adopted for the  TEGR charged field equations.  By a general gauge potential,   analytic solutions are derived. The  black hole singularities are investigated in \S IV. In \S V, the  energy conditions are discussed. In \S VI, the conserved charges are obtained. It is  shown that they are   divergent in the time direction. In \S VII, the method of ``regularization through relocalization''  is used  and finite conserved charges are obtained. Summary of the results are discussed in final section \S VIII.

\newsection{ A summary of  teleparallel equivalent of general relativity }
In the framework of  TEGR theory, the basic field
variables responsible for gravity are the tetrad
fields ${b_i}^\mu$ \cite{Wr}. The teleparallel condition leads to:  \begin{equation} \label{tp} b^i{}_{\mu;\nu}=
b^i{}_{\mu,\nu}-{W^\lambda}_{\mu \nu} {b^i}_\lambda=0, \end{equation} where $,$ and $;$ are the ordinary derivative and the covariant derivative respectively.  ${W^\lambda}_{\mu \nu}$, is  the non-symmetric Weitzenb\"ock connection.
${W^\mu}_{\lambda \nu}$ is  a non-symmetric connection  and it is defined as:  \begin{equation} {W^\lambda}_{\mu \nu}:=
{b_i}^\lambda {b^i}_{\mu, \nu}.\end{equation} Using  Eq.  (\ref{tp}) one can show that
 the curvature tensor vanishes identically. In  TEGR theory,  one can define the  spacetime metric,  $g_{\mu \nu}$,
 as
\begin{equation} g_{\mu \nu}:=  \lambda_{i j} {b^i}_\mu {b^j}_\nu, \end{equation}
with $\lambda_{i j}=(-1,+1,+1,+1)$ being the Minkowski
spacetime. A main property of TEGR is that one can  relate to any tetrad field
${b_i}^\mu$  a unique  metric   while, for a given metric $g_{\mu \nu}$,  one can connect  many tetrad fields due to the local Lorentz transformations.

The torsion  and the contortion tensors are defined as: \begin{eqnarray}
{T^\alpha}_{\mu \nu}  &:= &
{W^\alpha}_{\nu \mu}-{W^\alpha}_{\mu \nu} ={b_a}^\alpha
\left({b^a}{}_{\nu,\mu}-{b^a}{}_{\mu,\nu}\right),\nonumber\\
{K^{\mu \nu}}_\alpha &:= &
-\frac{1}{2}\left({T^{\mu \nu}}_\alpha-{T^{\nu
\mu}}_\alpha-{T_\alpha}^{\mu \nu}\right), \end{eqnarray}   where the contortion tensor
 can be rewritten in terms of connections as  ${K^{\mu}}_{\nu \rho}= {W^\mu}_{\nu \rho
}-\left \{_{\nu  \rho}^\mu\right\}$, with $\left \{_{\nu  \rho}^\mu\right\}$ being the Levi-Civita connection. The super-potential tensor ${S_\alpha}^{\mu \nu}$ is defined as: \begin{equation} {S_\alpha}^{\mu \nu}
:= \frac{1}{2}\left({K^{\mu
\nu}}_\alpha+\delta^\mu_\alpha{T^{\beta
\nu}}_\beta-\delta^\nu_\alpha{T^{\beta \mu}}_\beta\right), \end{equation}  and the torsion scalar takes the form \begin{equation} \label{ts} T :={T^\alpha}_{\mu \nu}
{S_\alpha}^{\mu \nu}. \end{equation} The gravitational action of  TEGR,  involving  the cosmological constant,  is defined as:
\begin{equation}
\label{lg1}  {\cal L}({b^i}_\mu)=\int
d^4x\; b\;\left[\frac{1}{16\pi}(T-2\Lambda)+{\cal L}_{ em}\right],
\end{equation}
 where $b:=\sqrt{-g}=det\left({b^i}_\mu\right),$
$ {\cal L}_{ em}$ is   the  Lagrangian  of   the  electromagnetic  field  and
$\Lambda$ is   the  cosmological constant.
We assume natural units where the gravitational  constant and the speed of light are $G = c = 1$. The Lagrangian of the electromagnetic field is defined as
\begin{equation}
{\cal L}_{
em}:=-\frac{1}{2}{ F}\wedge ^{\star}{F},
\end{equation}
where $F$ is the electromagnetic strength field which is defined as
 \begin{equation}
 F:= dA,
 \end{equation}
  with $A=A_{\mu}dx^\mu$ being the electromagnetic gauge
potential 1-form \cite{CGSV13}.  Carrying  out the variation of  Lagrangian (\ref{lg1})   with respect to the tetrad field ${b^i}_\mu$, one
 obtains the following field equations of   TEGR: \begin{equation} \label{fe}
{I^\nu}_{ \mu}\equiv e^{-1}{e^i}_\mu\partial_\rho\left(e{e_i}^\alpha
{S_\alpha}^{\rho \nu}\right)-{T^\alpha}_{\lambda \mu}{S_\alpha}^{\nu
\lambda}-\frac{1}{4}\delta^\nu_\mu(T-2\Lambda)+4\pi{{ \Theta}^\nu}_{ \mu}=0,\end{equation} \begin{equation} \label{fe1}
\partial_\nu \left( \sqrt{-g} F^{\mu \nu} \right)=0, \end{equation}
 where
 \[ {{ \Theta}^\nu}_{ \mu}=g_{\rho
\sigma}F^{\nu \rho}{F_\mu}^{\sigma}-\displaystyle{1 \over 4} {\delta_\mu}^{\nu} g^{\lambda \rho} g^{\epsilon \sigma} F_{\lambda \epsilon}
F_{\rho \sigma}, \] is the energy momentum tensor of the
electromagnetic field. In the following section we are going to apply the field equations (\ref{fe}) and (\ref{fe1}) to a tetrad field with  cylindrical symmetry.
\section{Exact  solutions with electromagnetic fields }\label{S3}
Now we are going to apply the  charged field equations of TEGR, Eqs. (\ref{fe}) and (\ref{fe1}), to the flat horizon   spacetime, which directly gives rise to the following  vierbein,  written in terms of cylindrical  coordinates ($t$, $r$, $\phi$, $z$) (see also  \cite{CGSV13}):
\begin{equation}\label{tetrad}
\hspace{-0.3cm}\begin{tabular}{l}
  $\left({b_{i}}^{\mu}\right)=\left( \sqrt{A(r)}, \; \frac{1}{\sqrt{A_1(r)}}, \; r, \; r\right)$,
\end{tabular}
\end{equation}
where $A(r)$ and $A_1(r)$ are two unknown functions of the radial coordinate $r$. Substituting  Eq. (\ref{tetrad}) into Eq. (\ref{ts}),  we evaluate the torsion scalar as\footnote{For the sake of simplicity,  we will write $A(r)\equiv A$,  \ \ $A_1(r)\equiv A_1$, \ \ $A'\equiv\frac{dA}{dr}$, $A'_1\equiv\frac{dA_1}{dr}$ $A''\equiv\frac{d^2A}{dr^2}$ and  $A''_1\equiv\frac{d^2A_1}{dr^2}$.}
\begin{eqnarray}\label{ts1}
\A \A T=2\frac{A'A_1}{rA}+2\frac{A_1}{r^2},
\end{eqnarray}
Applying Eq. (\ref{tetrad}) to the field equation (\ref{fe}) we get the following non-vanishing components:

\begin{eqnarray} \label{df1}
& & I_{t t}\equiv \frac{A}{r^4}\Biggl\{r^2A_1(a_\phi[2 b'_3-a_\phi]+a_{1z}[2 s'_1-a_{1z}])+s_\phi[ 2b_{z}-s_\phi]+b_{z}{}^2-r^2[A_1(s'_1{}^2+b'_1{}^2)+A_1\nonu
&& +rA'_1+r^2\Lambda]\Biggr\}=0,\nonumber\\
& & I_{r r}\equiv \frac{1}{r^4AA_1}\Biggl\{r^2AA_1(a_\phi[2 b'_3-a_\phi]+a_{1z}[2 s'_1-a_{1z}])+A s_\phi[ s_\phi-2b_{z}]+As_{1z}{}^2-r^2[AA_1(s'_1{}^2+b'_1{}^2)\nonu
&& -A_1A'r-A(A_1+r^2\Lambda)]\Biggr\}=0,\nonumber\\
&& I_{r \phi}\equiv I_{\phi r}=\frac{2(a_{1z}-s'_1)(s_\phi-b_{z})}{r^2}=0, \qquad \qquad I_{r z}=I_{zr}=\frac{2(a_{\phi}-b'_1)(s_\phi-b_{z})}{r^2}=0,\nonu
&& I_{\phi z}=I_{z \phi}=(a_{\phi}-b'_1)(s'_1-a_{1z})=0,\nonumber\\
&& I_{\phi \phi}\equiv \frac{1}{4r^2A^2}\Biggl\{2r^4AA_1A''-r^4A_1A'^2 +r^3AA'[rA'_1+2A_1]+2A^2\Biggl[2r^2A_1a_\phi[2 b'_1-a_\phi]-2s_\phi[s_\phi-2 b_{z}]\nonumber\\
&&+r^3A'_1-2r^2A_1a_{1z}(2s'_1-a_{1z})-2b_{z}{}^2-2r^2(b'_1{}^2A_1-r^2\Lambda-A_1s'_1{}^2)\Biggr] \Biggr\}=0,\nonumber\\
%
%
&& I_{z z}\equiv \frac{1}{4r^2A^2}\Biggl\{2r^4AA_1A''-r^4A_1A'^2 +r^3AA'[rA'_1+2A_1]+2A^2\Biggl[2r^2A_1a_\phi[a_\phi-2 b'_3]-2s_\phi[s_\phi-2 a_{1z}]\nonumber\\
&&+r^3A'_1+2r^2A_1a_{1z}(2s'_1-a_{1z})-2b_{1z}{}^2+2r^2(b'_1{}^2A_1+r^2\Lambda-A_1s'_1{}^2)\Biggr] \Biggr\}=0,\nonumber\\
\end{eqnarray}
where $a_\phi=\frac{da(\phi)}{d\phi}$, $a_{1z}=\frac{da_1(z)}{dz}$,  $b'_1=\frac{db_1(r)}{dr}$, $b_z=\frac{db(z)}{dz}$, $s'_1=\frac{ds_1(r)}{dr}$, $s_\phi=\frac{ds(\phi)}{d\phi}$ and  $a(\phi)$, $a_1(z)$,  $b(z)$, $b_1(r)$, $s_1(r)$ and $s(\phi)$   are the magnetic field strengths given by the general gauge potential as
\be \label{p} v :=[a(\phi)+a_1(z)]dr+[b(z)+b_1(r)]d\phi+[s(\phi)+s_1(r)]dz.\ee
The general   solutions  of the non-linear differential Eqs.  (\ref{df1}) have the form:
\begin{eqnarray} \label{sol}
& & i)\; A(r)=\frac{1}{A_1(r)}=\left(\frac{\Lambda r^3-3c_1}{3r}\right), \quad a(\phi)=c_2\phi,\quad a_1(z)=c_3z,\quad s(\phi)=c_4\phi,\quad b(z)=c_5z,\nonu
& & s_1(r)=c_6r,\quad b_1(r)=c_7r,\nonu
& &ii)\;  A(r)=\frac{1}{A_1(r)}=\left(\frac{\Lambda r^4-3c_1r-3c_8}{3r^2}\right), \quad a(\phi)=c_2\phi,\quad a_1(z)=c_3z,\quad s(\phi)=\pm \wp\phi, \nonu
& & b(z)=\wp^2 z, \quad s_1(r)=c_6r,\quad b_1(r)=c_7r, \qquad \wp=\frac{1\pm\sqrt{1\pm4\sqrt{c_8}}}{2},
\end{eqnarray}
where $c_i$,  $i=1\cdots 8$ are constants of integration. Eqs. (\ref{sol}) shows that when the constant $c_8=0$ then the second set will be identical to the first set and the constant $\wp$, after some re-scaling, can  be related to the constants $c_4$ and $c_5$ of the first set.

\section{The physical properties of  solutions}

The metric of solutions (\ref{sol})  take the form
\begin{eqnarray}  \label{me}
&&ds^2{}_1=-\left(\frac{\Lambda r^3-3c_1}{3r}\right) dt^2+\frac{dr^2}{\left(\frac{\Lambda r^3-3c_1}{3r}\right)}+r^2(d\phi^2+dz^2)\;,\nonu
& &ds^2{}_2=-\left(\frac{\Lambda r^4-3c_1r-3c_8}{3r^2}\right)dt^2+\frac{dr^2}{\left(\frac{\Lambda r^4-3c_1r-3c_8}{3r^2}\right)}+r^2(d\phi^2+dz^2)\;. \end{eqnarray}
Eqs. (\ref{me}) show that the metrics asymptotically behave as AdS/dS spacetime. Furthermore Eqs. (\ref{me}) show that the first metric is  static  without any charge. The second metric has a charge which comes from the term of order $O(\frac{1}{r^2})$.   The second Eq.  (\ref{me}) can be rewritten as
\begin{eqnarray}  \label{me1}
ds^2{}_2=-\left(\frac{\Lambda r^2}{3}-\frac{m}{r}-\frac{q^2}{r^2}\right)dt^2+\left(\frac{\Lambda r^2}{3}-\frac{m}{r}-\frac{q_m{}^2}{r^2}\right)^{-1}dr^2+r^2(d\phi^2+dz^2)\;, \end{eqnarray} where $m=c_1$ and $q_m=\sqrt{c_m}$. The metric (\ref{me1}) is similar to  the AdS/dS Reissner-Nordstr\"om solution \cite{NC03}. Here we want to stress   that the source of term  ${\cal O}(r^{-2})$ in metric (\ref{me1}) comes from the presence of magnetic field while,  in the  Reissner-Nordstr\"om case,  such a term is related to  the source of electric field.

Inserting Eqs. (\ref{sol}) into Eq. (\ref{ts}) we get
\begin{equation} \label{tss}
 T=2\Lambda,  \qquad\qquad
T= \frac{2({c_8}+\Lambda r^4)}{r^4}, \end{equation}
   which shows that the scalar torsion is not constant in the second case.  From Eqs. (\ref{tss}), it is easy to see  that the torsion scalar of second case reduces to the first case as soon as  the constant $c_8=0$. The behavior of the scalar torsion is given in Fig. 1.
   \begin{figure}
\centering
\includegraphics[scale=.3]{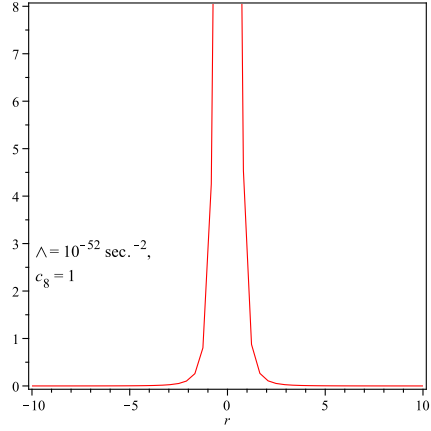}
\caption{The behavior of the torsion scalar for the second  solution (\ref{sol}).}
\label{fig. 1}
\end{figure}

Now we are going  to calculate the  {\it singularities of solutions (\ref{sol})}.  The first step to discuss this issue is to find at which value of $r$  the functions $A(r)$ and $A_1(r)$ become zero or infinity.  The curvature and torsion invariants that arise from the first  solution (\ref{sol}),
using the Levi-Civita and Weitzenb\"ock connections,  take the form:
\begin{eqnarray} R^{\mu \nu \lambda \rho}R_{\mu \nu \lambda \rho} \A = \A \frac{4(2\Lambda^2r^6+9c_1{}^2)}{3r^6},
 \qquad
R^{\mu \nu}R_{\mu \nu} = 4\Lambda^2,\qquad
R =-4\Lambda,\nonumber\\
T^{\mu \nu \lambda}T_{\mu \nu \lambda} \A=\A \frac{4\Lambda^2r^6-12\Lambda c_1r^3+27{c_1}^2}{2r^3(3{c_1}-\Lambda r^3)}, \qquad
T^\mu T_\mu = \frac{3(3{c_1}-2\Lambda r^3)^2}{4r^3(3{c_1}-\Lambda r^3)},\quad
T(r)=-2\Lambda, \nonumber\\
&&  \nabla_\alpha T^\alpha=3\Lambda, \qquad \qquad \Rightarrow R=-T-2\nabla_\alpha T^\alpha.
\end{eqnarray}
and for the second solution,  we get the invariants
 \begin{eqnarray} R^{\mu \nu \lambda \rho}R_{\mu \nu \lambda \rho} \A = \A \frac{4(2\Lambda^2r^8+6c_8[6c_1+7{c_8}r]+9r^2{c_1}^2)}{3r^8},
 \quad
R^{\mu \nu}R_{\mu \nu} = \frac{4({c_8}^2+\Lambda^2r^8)}{r^8},\quad
R =-4\Lambda,\nonumber\\
T^{\mu \nu \lambda}T_{\mu \nu \lambda} \A=\A \frac{4\Lambda^2r^8-8\Lambda c_8r^4-12\Lambda{c_1}r^5+27r^2{c_1}^2+60{c_1}c_8r+36{c_8}^2}{2r^4(3{c_1}r-\Lambda r^4+3c_8)}, \nonumber\\
T^\mu T_\mu \A=\A \frac{3(3{c_1}r-2\Lambda r^4+2{c_8})^2}{4r^4(3{c_1}r-\Lambda r^4+3c_8)},\quad
T(r)=-\frac{2(\Lambda r^4+{c_8})}{r^4},\qquad \nabla_\alpha T^\alpha=\frac{(3\Lambda r^4+{c_8})}{r^4}, \nonumber\\
&&  \Rightarrow R=-T-2\nabla_\alpha T^\alpha.
\end{eqnarray}
 The above calculations show that:\vspace{0.1cm}\\
 a)-Except for the scalars $R^{\mu \nu}R_{\mu \nu}$, $R$, $ \nabla_\alpha T^\alpha$,  $T$ of the first solution and $R$ of the second solution,  all the above invariants show infinite behavior at $r=0$ which represents a true singularity.\vspace{0.1cm}\\
  b)- For $c_1=\frac{r^3\Lambda}{3}$ for the first solution,  we get the horizon on the metric. For this value the curvature invariants are finite but torsion invariants diverge, i.e.
   \[T^{\mu \nu \lambda}T_{\mu \nu \lambda}\rightarrow \infty, \qquad \textrm{and} \qquad  T^{\mu }T_{\mu} \rightarrow \infty.\]  This means that, on the horizon, the torsion invariants  diverge. The reason that leads the curvature invariants  to have finite value but torsion invariants diverge is the local Lorentz transformations. This can be seen clearly  from the calculations of the scalar torsion $T(r)$ which is finite on the horizon due to its invariant under local Lorentz transformations however, the scalars  $T^{\mu \nu \lambda}T_{\mu \nu \lambda}$ and  $T^\mu T_\mu$ are not finite because they are not invariant under local Lorentz transformations. The same discussion can be applied when $c_8=\frac{r^3\Lambda -3c_1}{3}$ for the second set of solution (\ref{sol}).   \vspace{0.1cm}\\
 c)- The horizons of solutions (\ref{sol})  are respectively given for $c_1=\frac{r^3\Lambda}{3}$  and $c_8=\frac{r^3\Lambda -3c_1}{3}$.

 Let us  now  discuss some thermodynamical quantities related to the solution  (\ref{me1}). To this aim,  we calculate the horizons of the function \begin{equation}{\cal N}=\frac{\Lambda r^2}{3}-\frac{m}{r}-\frac{q^2}{r^2}.\end{equation} The above equation has 4 roots,  3 of them are imaginary  while the fourth one is real and takes the form
 \begin{equation}\frac{3^{2/3}[2^{5/6}\{X^{2/3}-4q^2 \Lambda_1{}^{1/3}\}^{3/4}+2^{7/12}\sqrt{X^{2/3}\sqrt{2(X^{2/3}-4q^2 \Lambda_1{}^{1/3})}+ 2^{5/2}  \Lambda_1{}^{1/3}q^2(X^{2/3}-4q^2\Lambda_1^{1/3})-12m\sqrt{X}})]}{12\Lambda^{1/3}X^{1/6}[X^{2/3}-4q^2\Lambda_1{}^{1/3}]^{1/4}},\end{equation}  where $\Lambda_1=12\Lambda$ and $X=9m^2+\sqrt{3(256q^2\Lambda+27m^4)}$.  To ensure we have a real root, we must have ${\displaystyle \Lambda>-\frac{27m^4}{256 q^6}}$. The behavior of the horizon is drawn in Figure 2 which shows that we have only one horizon.
 \begin{figure}
\centering
\includegraphics[scale=.3]{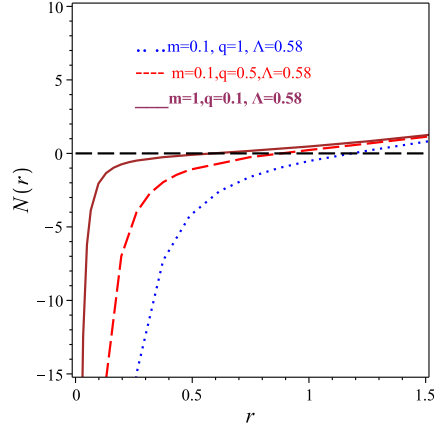}
\caption{The horizon of   solution (\ref{me1}).}
\label{fig2}
\end{figure}
 The Hawking temperature is defined as \cite{NO13}
  \begin{equation}
T_h = \frac{{\cal N}'(r_h)}{4\pi},
\end{equation}
where the event horizon is located at $r = r_h$ which is the largest positive root of ${\cal N}(r_h) = 0$ that fulfills the condition ${\cal N}'(r_h)\neq 0$.
The Hawking temperatures associated with the black hole solution (\ref{me1}) is calculated as
\begin{eqnarray} \label{m44}
{T_h}=\frac{3r_h{}^4\Lambda+q^2}{4\pi r_h{}^3},
\end{eqnarray}
where ${T_h}$ is the Hawking temperature at the event horizon. We represent the Hawking temperature in Figure 3.  This last figure   shows that the temperature  is always positive.
 \begin{figure}
\centering
\includegraphics[scale=.3]{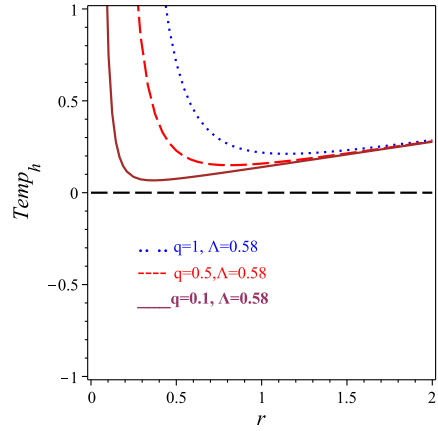}
\caption{The Hawking temperature of   solution (\ref{me1}).}
\label{fig3}
\end{figure}

 \newsection{Energy conditions}
 An important issue  is related to the possible violation of the energy conditions in cosmology or strong field regime. In GR, there are four types of energy conditions known as:
The strong energy condition (SEC), the null energy condition (NEC), the dominant energy condition (DEC) and the weak energy condition (WEC) \cite{HE,Nepjp,C4,Nahep}. The SEC and the NEC arise  from the
 structure of the gravitational field  related to  the dynamics of  matter. It is related to the   Raychaudhuri
equation which leads the time
  expansion of the scalar $\theta$ in terms of
quantities like the Ricci tensor,  the shear tensor $\sigma^{\mu \nu}$ and the rotation
$\omega^{\mu \nu}$ for both  time and light-like curves. These relations have the form:
\ba
 \frac{d\theta}{d\tau}=-\frac{1}{3}\theta^2-\sigma_{\mu \nu}\sigma^{\mu \nu}+\omega_{\mu \nu}\omega^{\mu \nu}-R_{\mu \nu} u^\mu u^\nu=-\frac{1}{3}\theta^2-\sigma_{\mu \nu}\sigma^{\mu \nu}+\omega_{\mu \nu}\omega^{\mu \nu}-R_{\mu \nu} k^\mu k^\nu,\ea
 where $u^\mu$  is an arbitrary time-like
vector and $k^\mu$ is an arbitrary null vector.
As a consequence  of the attraction,  one can show
\be \label{rc} R_{\mu \nu} u^\mu u^\nu \geq 0, \qquad \qquad \qquad \qquad R_{\mu \nu} k^\mu k^\nu\geq 0.\ee Eqs. (\ref{rc}) can be rewritten as
\be \label{se} R_{\mu \nu} u^\mu u^\nu=\left({\cal T}_{\mu \nu}-\frac{\cal T}{2}g_{\mu \nu}\right)u^\mu u^\nu \geq 0, \qquad   R_{\mu \nu} k^\mu k^\nu=\left({\cal T}_{\mu \nu}-\frac{\cal T}{2}g_{\mu \nu}\right)k^\mu k^\nu\geq 0,\ee
which are  the SEC and the NEC, respectively for a given source of matter ${\cal T}_{\mu\nu}$.
In the case of perfect-fluid matter, the SEC and NEC given
by (\ref{se})  impose the following constraints $\rho+3p\geq 0$ and $\rho+p\geq 0$ to be satisfied, while the WEC and DEC
require the conditions $\rho\geq 0$ and $\rho\pm p\geq 0$, respectively for
consistency.

The energy-momentum components of the first  solution (\ref{sol}) are vanishing. This means that the first solution (\ref{sol}) is a vacuum solution.   The non-vanishing components of the energy-momentum tensor of the second solution  Eq. (\ref{sol}) have the form
\be \label{st1}
{\cal T}^0{}_0=-{\cal T}^3{}_3={\cal T}^1{}_1=-{\cal T}^2{}_2=-\frac{c_8}{2r^4}.\ee
Eqs. (\ref{st1}) show that the WEC is violated unless $c_8<0$ and the DEC is satisfied for  $\rho-p\geq 0$. However  the DEC, NEC and SEC are all satisfied.
\section{Einstein-Cartan theory and conserved currents}
The above considerations can be extended in the framework og the Einstein-Cartan theory. Let us define the
 Einstein-Cartan Lagrangian as  \cite{OR6}:
 \be \label{lg}
{\cal L}(\vartheta^i, \
{\Gamma^j}_{k})=-\frac{1}{2\kappa}\left(R^{i j}\wedge
\eta_{i j}-2\Lambda \eta\right),\ee where $\vartheta^i$ is the co-frame, ${\Gamma^j}_{k}$ is the connection one-form and
$\kappa$ is the gravitational coupling constant that now we explicitly redefine. Lagrangian (\ref{lg}) is invariant under diffeomorphism and local Lorentz transformations \cite{OR6}.  The variation of Eq. (\ref{lg})  leads to the
canonical energy-momentum and rotational  gauge field momentum
with  the forms \cite{OR6,OR7} \ba && E_{i}:= -\frac{1}{2\kappa}\left(R^{ j k}\wedge
\eta_{i j k}-2\Lambda \eta_i\right) ,   \qquad \qquad  H_{i j}:=\frac{1}{2\kappa}\eta_{i j},\ea
with $\eta_{i j}$ being a 2-form defined in Appendix A and $R^{ j k}$ is the curvature 2-form. The conserved quantity of the
gravitational field of (\ref{lg}) is  \cite{OR6}
\be \label{ch}  \jmath[\xi]=\frac{1}{2\kappa}d\left\{^{^{*}}\left[dk+\xi\rfloor\left(\vartheta^i\wedge
T_j\right)\right]\right\},\ee
 where $ k=\xi_i\vartheta^i,$ and  $ \xi^i=\xi\rfloor\vartheta^i $.
Here $*$ denotes the Hodge duality, $\xi$ is an arbitrary vector
field $\xi=\xi^i\partial_i$ and $\xi^i$ are four  parameters $\xi^0$, $\xi^1$, $\xi^2$ and $\xi^3$.
We are working in TEGR theory which is equivalent to GR, therefore the torsion is vanishing
  and  the total charge, given by Eq. (\ref{ch}),  takes the form
\be \label{ch1} {{\cal Q}}[\xi]=\frac{1}{2\kappa}\int_{\partial S}{^*}dk. \ee
This invariant conserved quantity ${{\cal Q}}[\xi]$ was   previously defined  by   Komar  \cite{Ka2}--\cite{Ka3}. The quantity ${{\cal Q}}[\xi]$ is conserved  and invariant
(for any given vector field $\xi$) under  general coordinate transformations.

The coframe $\vartheta^{\delta}$ of
solution (\ref{sol}), using tetrad (\ref{tetrad}) , has the form: \ba \label{co} {\vartheta}^{{0}} =\sqrt{A(r)}dt,\qquad
{\vartheta}^{{1}}=\frac{1}{\sqrt{A_1(r)}}dt, \qquad {\vartheta}^{{2}}= r d\phi, \qquad
 {\vartheta}^{{3}}=rdz.  \ea
Using Eqs.
(\ref{co}) into Eq. (\ref{ch}), we get \be \label{k1}  k=A(r)\xi_0dt - \frac{\xi_1dr}{A_1(r)}-r^2\xi_2d\phi -r^2\xi_3dz.\ee After some algebra, the total derivative of Eq. (\ref{k1}) has the
form \be \label{dk} dk= A'(r)\xi_0(dr \wedge dt)+2r\xi_2(d\phi \wedge dr)+2r\xi_3(dz \wedge dr).
\ee
 Using the inverse of Eq. (\ref{k1}), (i.e. we write $dt$, $dr$, $d\theta$ and $d\phi$ in terms of ${\vartheta}^{\hat{0}}$, ${\vartheta}^{\hat{1}}$, ${\vartheta}^{\hat{2}}$ and ${\vartheta}^{\hat{3}}$) and
    substituting Eq. (\ref{dk}) in Eq. (\ref{ch1}) and applying
the Hodge-dual to $dk$,  we finally get the total conserved charge in
the form
\be \label{ch2} {{\cal Q}}[\xi_t]=\frac{\xi_0(2r^3\Lambda+3c_1)}{6}, \qquad \qquad {{\cal Q}}[\xi_r]={{\cal Q}}[\xi_\theta]={{\cal Q}}[\xi_\phi]=0,\ee
Using the same algorithm for the second  solution (\ref{sol}),  we get
\be \label{cc2} {{\cal Q}}[\xi_t]=\frac{\xi_0(2r^4\Lambda+3c_1r+6c_7)}{6r}, \qquad \qquad {{\cal Q}}[\xi_r]={{\cal Q}}[\xi_\theta]={{\cal Q}}[\xi_\phi]=0,\ee
  Eqs. (\ref{ch2}) and (\ref{cc2}) show that the total conserved charges of
solutions (\ref{sol}), using tetrad (\ref{tetrad}) and Eq. (\ref{ch1}),  are  divergent when $r\rightarrow \infty$.
Therefore, Eq. (\ref{ch1}) needs a regularization.
\section{Regularization via relocalization}
The conserved quantity given by Eq. (\ref{ch1}) is invariant under diffeomorphism   and local
Lorentz transformations. Besides these transformations
there is another  issue in the definition of the conserved
quantities which lies in the fact that the field equations
 allow for a relocalization of the gravitational field
momenta \cite{OR6}. Thus, the conserved currents can be altered  through the relocalization of
translational and rotational momenta. A relocalization  generated by
 altering  the Lagrangian  of the gravitational field by a total
derivative is  given by \be {\cal L}'={\cal L}+d\aleph, \qquad \textrm{where} \qquad
\aleph=\aleph({{{\vartheta^{{}^{{}^{\!\!\!\!}}}}{_{}{_{}{_{}}}}}^{i}},
{\Gamma_i}^j, T^i, {R_i}^i).\ee  The second term exists in the Lagrangian, i.e.,
$d\aleph$ modifies only the boundary part of the action, allowing the field equations to be invariant \cite{OR6}. It is straightforward that the total conserved
quantities can be regularized by means of a relocalization of the gravitational field momenta.
It is shown that the most accurate method, that can solve the strange result
derived in Eqs. (\ref{ch2}) and (\ref{cc2}), is to use  relocalization which is originated by a boundary term in the Lagrangian. Here we use the relocalization
\[ H_{i j}\rightarrow H'_{i j}=H_{i j}-2\alpha\eta_{i j k l}R^{k l},\] which  is originated by altering    the Lagrangian as \cite{OR6}
\[{\cal L}\rightarrow {\cal L}'={\cal L}+\alpha d\aleph,\] where \[ H'_{i j}=\left(\frac{1}{2\kappa}-\frac{4\alpha \Lambda}{3}\right)\eta_{i j}-2\alpha \eta_{i j k l}\left(R^{k l}-\frac{\Lambda}{3}{\vartheta}^{k}{\vartheta}^{l}\right).\] We assume $\alpha$, that appears in the above equation to have the form $\frac{3}{8\Lambda \kappa}$ to insure the removal of the divergence that appear in Eqs.  (\ref{ch2}) and (\ref{cc2}). Therefore,  the conserved
charge, using the relocalization method, takes the form
\be \label{ch4} {{\cal J}}[\xi]=-\frac{3}{4\kappa \Lambda }\int_{\partial S}
\eta_{i j k l}\Xi^{i j} W^{k l}, \ee where
$W^{i j}$ is the Weyl 2-form defined by \be W^{i j}=\frac{1}{2}{C_{k l}}^{i j}{\vartheta}^{k}\wedge {\vartheta}^{l},\ee with ${C_{i j}}^{k l}={b_i}^\mu {b_j}^\nu {b^k}_\alpha {b^l}_\beta {C_{\mu \nu}}^{\alpha \beta}$ being  the Weyl tensor and  $\Xi^{i j}$
defined as\footnote{The detailed derivation of Eq. (\ref{ch4}) is found in references
 \cite{OR6,OR7,OR8}.} \be \Xi_{i j}:=\frac{1}{2}e_j\rfloor
e_i \rfloor dk.\ee  The conserved
currents ${{\cal J}}[\xi]$  are invariant under both coordinate and local Lorentz transformations. These
currents ${{\cal J}}[\xi]$ are related to a given  vector field $\xi$ on the
spacetime of the manifold.

 We calculate the necessary components needed for Eq. (\ref{ch4}). The non-vanishing components of $\Xi^{i j}$ have the form\footnote{The non-vanishing components of
 Weyl tensor are given in Appendix B.}
\ba \Xi_{01} \A = \A-\frac{\xi_0(2\Lambda r^3+3c_1)}{6r^2}, \qquad  \Xi_{13}= \frac{\xi_3(3c_1-\Lambda r^3)}{\sqrt{3r}}. \ea
  Using Eqs. (\ref{ch4}), we get \be\label{ch5} \eta_{i j k l}\Xi^{i j}
W^{k l}=\frac{ 2c_1\xi_0(2\Lambda r^3+3c_1)(dz \wedge d\phi)}{3r^3}\ee Substituting Eq. (\ref{ch5}) in
(\ref{ch4}) we finally get \be \label{ch6} {{\cal
J}}[\xi_t]=\frac{c_1}{2},\qquad {{\cal
J}}[\xi_r]={{\cal
J}}[\xi_\theta]={{\cal
J}}[\xi_\phi]=0.\ee Eqs. (\ref{ch6}) show that the constant $c_1$ may take the value  $c_1=\frac{M}{2}$  such that the total mass of Eqs. (\ref{ch6}) takes the form \cite{LZ, Dm}
\be E=M+\left(\frac{1}{r}\right).\ee
By the same method, we can get the conserved charge of the second solution (\ref{sol}). It has the form
\be E=M+\left(\frac{c_8}{r}\right)+O\left(\frac{1}{r^2}\right),\ee which shows that the constant $c_8$ behaves as the electric charge.
\newsection{ Discussion and conclusions }

 Including a  magnetic field in the metric is a challenging  issue to get exact solutions in theories of gravity. Despite of this difficulty, some analytic solutions
have been derived, like  \cite{ M64,B54},  where  a magnetic "universe", including a magnetic field in
the $z$ direction, is considered.  Furthermore Gutsunaev  and Man'ko found a solution  where a magnetic dipole is present \cite{GM88}. Of course it is always possible to study an arbitrary shape for the magnetic field and solve the resulting Einstein equations. In this study, we have addressed the problem of deriving charged black hole solutions, in TEGR theory, involving cosmological constant and  using a general gauge potential including magnetic fields only. For this purpose, we have applied a tetrad field with two unknown functions and assumed a cylindrical symmetry for the charged field equations of TEGR. We have used a gauge potential which contains 6 unknown functions. Finally, we obtained a system of nonlinear differential equations that has been solved exactly. The solution of this system has two cases: In the first one, the torsion scalar has a constant value and all the components of the energy momentum, which depends on the charge fields,  are  identically vanishing  while all the components of the magnetic field have non-trivial values. The second case  contains an   integration constant  which gives a nontrivial value to the torsion scalar which  becomes trivial   when this constant is equal zero. It is worth noticing   that this constant is related to some component of the magnetic field.

We then discuss the energy conditions related to these solutions and show that the first set satisfies these conditions because it has a trivial value of the energy momentum tensor. However, for the second set, the energy conditions are satisfied under certain constraints. We have also  discussed the singularities of the two sets and have discussed  the horizons of each set. Finally, we have calculated the conserved quantities related  to each set and have shown that the Komar formula gives a divergent quantity on the temporal components.

Therefore, we have applied the regularization through relocalization in order to calculate the conserved quantities. For the first set, we have shown that the only conserved quantity is the energy and have related the constant that appeared in the calculation of energy to the ADM mass. The conservation of the second set gives, besides the ADM mass, another term which is related to the constant that makes the torsion scalar a dynamical one. So we can explain the contribution of this constant as related to the magnetic field. The most interesting thing is that the sign of this term  is different from the sign of the term of  Reissner-Nordstr\"om spacetime whose source of charge comes from the electric charge \cite{NS}. In a forthcoming paper, we will extend these considerations to more general TEGR models like those discussed in \cite{cai}. 

It is important to stress that magnetic teleparallel solutions can be obtained also in spherical symmetry adopting    procedures similar to that considered   in this paper. In fact, it is easy to see that the assumption (\ref{tetrad}) can be recast for spherical coordinates considering suitable vierbien fields. However, the  magnetic field has to be adapted to the spherical symmetry and one obtains different forms for the  functions $A(r)$ and $A_1(r)$.

A final remark concerns possible astrophysical applications of the present results. As we said,  the value of the torsion scalar depends on the strength of the magnetic field and this fact could have observational consequences on magnetic astrophysical systems. As reported in \cite{andrade},  torsion plays a dynamical role on magnetic vortex line curves of magnetars. In particular, torsion contributes to the oscillations of the magnetar and to the equation of state of such systems. Furthermore, in \cite{lyutikov}, several observational evidences are reported for neutron star magnetospheres related to  binary pulsars, Crab pulses and magnetars. In all these cases, the strict relation between torsion and magnetic field  could contribute to figure out  the dynamics. A detailed analysis in this direction will be developed in a forthcoming study.

\appendix

\section{ Notation used in the calculations of conserved currents }

The   indices ${\it i, j, \cdots }$ are used for the (co)frame components while   $\alpha$, $\beta$,
$\cdots$ label the local holonomic
spacetime coordinates.  Exterior product is defined as $\wedge$, while the
interior  is denoted
by $\xi \rfloor \Psi$. The vector basis, dual to the  1-forms
$\vartheta^{i}$,  is denoted by $e_i$. They satisfy the condition
$e_i \rfloor \vartheta^{j}={\delta_i}^j$. Using the
local coordinates $x^\mu$, we have $\vartheta^{i}={b^i}_\mu
dx^\mu$ and $e_i={b_i}^\mu \partial_\mu$ where ${b^i}_\mu$ and
${b_i}^\mu $ are the covariant and contravariant components of
the tetrad field. The volume is defined as $\eta:=\vartheta^{\hat{0}}\wedge \vartheta^{\hat{1}}\wedge
\vartheta^{\hat{2}}\wedge\vartheta^{\hat{3}}$ which is a 4-form.   Moreover, by using the interior product one can  define
\[\eta_i:=e_i \rfloor \eta = \ \frac{1}{3!} \
\epsilon_{i j k l} \ \vartheta^j \wedge
\vartheta^k \wedge \vartheta^l,\]
where
$\epsilon_{ i j k l}$ is totally antisymmetric
with $\epsilon_{0123}=1$. \[\eta_{ i j}:=e_j \rfloor \eta_i =
\frac{1}{2!}\epsilon_{i j k l} \
\vartheta^k \wedge \vartheta^l,\qquad \qquad
\eta_{i j k}:=e_k
\rfloor \eta_{i j}= \frac{1}{1!} \epsilon_{i j k l} \ \vartheta^l,\]  that are the bases for 3-, 2-
and 1-forms respectively. Finally, \[\eta_{i j k l}
:=e_l \rfloor \eta_{i j k}=
e_l \rfloor e_k \rfloor e_j \rfloor e_i \rfloor
\eta,\] is the Levi-Civita tensor density. The $\eta$-forms
satisfy the useful identities: \ba \vartheta^i \wedge
\eta_j \A:=  \A \delta^i_j
\eta, \qquad \vartheta^i \wedge \eta_{j k}  := \delta^i_k \eta_j-\delta^i_j \eta_k,
\qquad  \vartheta^i \wedge \eta_{j k l} := \delta^i_j \eta_{k l}+\delta^i_k
\eta_{l j}+\delta^i_l \eta_{ j k}, \nonumber\\
\vartheta^i \wedge \eta_{j k l n}  \A :=\A \delta^i_n \eta_{j k l}-\delta^i_l \eta_{j k n
}+\delta^i_k \eta_{ j l n}-\delta^i_j
\eta_{k l n}. \ea
\section { Calculations of the Weyl and $W^{\mu \nu}$ tensors}

The non-vanishing components of Weyl
tensor, using solutions (\ref{sol}),  have the form:
\ba && C_{0101}=-C_{0110} =C_{1010}=-C_{1001}=2C_{0220}=-2C_{0202}=2C_{0330}=-2C_{0303}=-2C_{2020}=2C_{2002}\nonumber\\
 && =2C_{3003}=-2C_{3030}=
2C_{1212}=-2C_{1221}= 2C_{1313}=- 2C_{1331}=-2C_{2112}=2C_{2121}=-2C_{3113}\nonumber\\
 &&=2C_{3131}= -C_{2323}=C_{2332}=-C_{3232}=C_{3223}=-\frac{c_1}{r^3},
\ea
and the non-vanishing components of the tensor $W^{\mu \nu}$ take the form
\ba && W^{01}=\frac{c_1}{r^3}(dt \wedge dr),\qquad W^{02}=\frac{\sqrt{3c_1-\Lambda r^3}}{2\sqrt{3r^5}}(d\phi \wedge dt),\nonumber\\
&& W^{03}=\frac{\sqrt{3c_1-\Lambda r^3}}{2\sqrt{3r^5}}(dz \wedge dt),\qquad W^{12}=\frac{3c_1}{2\sqrt{3r^3(3c_1-\Lambda r^3)}}(d\phi \wedge dr),\nonumber\\
&&  W^{13}=\frac{3c_1}{2\sqrt{3r^3(3c_1-\Lambda r^3)}}(dr \wedge dz), \qquad W^{23}=\frac{c_1(d\phi \wedge dz)}{r}.\nonumber\\
\ea
\subsection*{Acknowledgments}
 This work is partially supported by the Egyptian Ministry of Scientific Research under project No. 24-2-12. S.C. acknowledges COST action CA15117 (CANTATA), supported by COST (European Cooperation in Science and Technology).
 
\end{document}